\documentclass[aps,prl,twocolumn,groupedaddress]{revtex4}
\usepackage[dvips]{graphics, color}
\usepackage{amsmath}
\usepackage{amssymb}
\usepackage{float}
\usepackage{bm,times}
\usepackage{graphicx}
\newcommand{\be}{\begin{equation}}
\newcommand{\ee}{\end{equation}}
\newcommand{\bea}{\begin{eqnarray}}
\newcommand{\eea}{\end{eqnarray}}

\begin{document}

\title{Measuring nanomechanical motion with a microwave cavity interferometer}

\author{C. A. Regal}
\email[Current address: Norman Bridge Laboratory of Physics 12-33, California Institute of Technology, Pasadena, California
91125]{}
\author{J. D. Teufel}
\author{K. W. Lehnert}
\email[E-mail: ]{konrad.lehnert@jila.colorado.edu}
\address{
JILA, National Institute of Standards and Technology and the University of Colorado, and Department of Physics University of
Colorado, Boulder, Colorado 80309, USA}

\date{April 6, 2008}

\begin{abstract}
In recent years microfabricated microwave cavities have been extremely successful in a wide variety of detector
applications.  In this article we focus this technology on the challenge of quantum-limited displacement detection of a
macroscopic object.  We measure the displacement of a nanomechanical beam by capacitively coupling its position to the
resonant frequency of a superconducting transmission-line microwave cavity.  With our device we realize near
state-of-the-art mechanical force sensitivity ($3 \, \rm{aN/\sqrt{Hz}}$) and thus add to only a handful of techniques able
to measure thermomechanical motion at 10's of milliKelvin temperatures.  Our measurement imprecision reaches a promising 30
times the expected imprecision at the standard quantum limit, and we quantify our ability to extract measurement backaction
from our results as well as elucidate the important steps that will be required to progress towards the full quantum limit
with this new detector.
\end{abstract}

\maketitle

The advent of micro and nanomechanical resonators has brought the long-standing goal of exploring quantum effects such as
superposition and entanglement of macroscopic objects closer to reality.  With this ability one could experimentally study
decoherence of superposition states thus elucidating questions about the interface between the quantum and classical worlds.
Micro and nanomechanical resonators have hastened progress towards macroscopic quantum limits by providing high-frequency,
small-dissipation, yet low-mass resonators. Still it remains a challenge to freeze out the thermomechanical motion of these
objects to leave only zero-point fluctuations $\delta x _{zp}$ and, equally importantly, to detect motion at this level. The
problem of detection at the quantum limit is in itself intriguing. As the imprecision of any detector is decreased,
measurement backaction emerges to enforce the Heisenberg constraint, which for continuous displacement detection is
$S_x^{im}S_F^{ba} \ge \hbar^2$.  Here $S_x^{im}$ and $S_F^{ba}$ are respectively the displacement imprecision and backaction
force spectral densities.  In fact, at the minimum allowed total position uncertainty, referred to as the standard quantum
limit (SQL), the measurement imprecision and backaction must together contribute an uncertainty equal to the zero-point
fluctuations.

A widely used displacement detector that, in principle, is capable of reaching the SQL is an optical cavity interferometer
with a moving mirror \cite{Caves1981a}.  Physically, at the SQL ones knowledge of the mirror position is limited equally by
shot noise in the output signal and by motion of the mirror due to quantum fluctuations in the intracavity radiation
pressure. This limit has long been of interest in quantum optics and in the gravitational wave detection community.  Optical
cavities generally outperform all other displacement detectors with regard to measurement imprecision; they can achieve
shot-noise limited position sensitivity as low as $\sim 10^{-19}$ $\rm m/\sqrt{Hz}$ \cite{Arcizet2006}.  Still reaching the
SQL has historically been a challenge due to the inaccessibility of quantum backaction effects
\cite{Tittonen1999a,Braginsky1985}. Recent experimental progress using low-mass mirrored microcantilevers has made radiation
pressure effects more observable \cite{Arcizet2006b,Gigan2006,Kleckner2006,Schliesser2006,Thompson2007}.

However, the most successful approaches to the SQL to date have been electromechanical experiments.  These experiments take
place ``on-chip" in a dilution refrigerator where thermomechanical motion is significantly reduced, and the mechanical
objects are typically nanoscale and hence even less massive than microcantilevers.  Examples include using a single electron
transistor \cite{LaHaye2004,Naik2006} or an atomic point contact \cite{FJ2007} for the displacement readout of a nanoscale
flexural beam.  Electromechanical experiments have observed a displacement uncertainty less than 10 times the total
uncertainty added by the measurement at the SQL and evidence for backaction \cite{LaHaye2004,Naik2006}. Still
electromechanical experiments have not achieved the full quantum limit typically due to technical noise sources common to
mesoscopic amplifiers.

In this article we present experiments that use the principles and advantages of an optical cavity interferometer with a
moving mirror yet employ ``light" at microwave frequencies.  Operating at microwave frequencies allows us to also benefit
from technology associated with electromechanical systems, such as low-mass mechanical objects and dilution refrigerator
temperatures. Specifically, we embed a nanomechanical flexural resonator inside a superconducting transmission-line
microwave cavity, where the mechanical resonator's position couples to the cavity capacitance and thus to the cavity
resonance frequency. Changes in this frequency can be sensitively monitored via homodyne detection of the phase shift of a
microwave probe signal. Advantages of superconducting transmission-line cavities include large demonstrated quality factors
($Q>10^5$) \cite{Day2003} and a tiny mode volume. Additionally the cavities are fabricated via a single deposition of a
thin, superconducting film and thus are scalable as well as compatible with patterning of other nanoscale devices.  These
advantages have been leveraged in an array of other recent applications including microwave kinetic inductance detectors
(MKIDS) \cite{Day2003}, achieving circuit QED \cite{Wallraff2004}, and readout of superconducting quantum interference
devices (SQUIDS) \cite{Mates2008}.

\begin{figure} \begin{center}
\includegraphics[width=220pt]{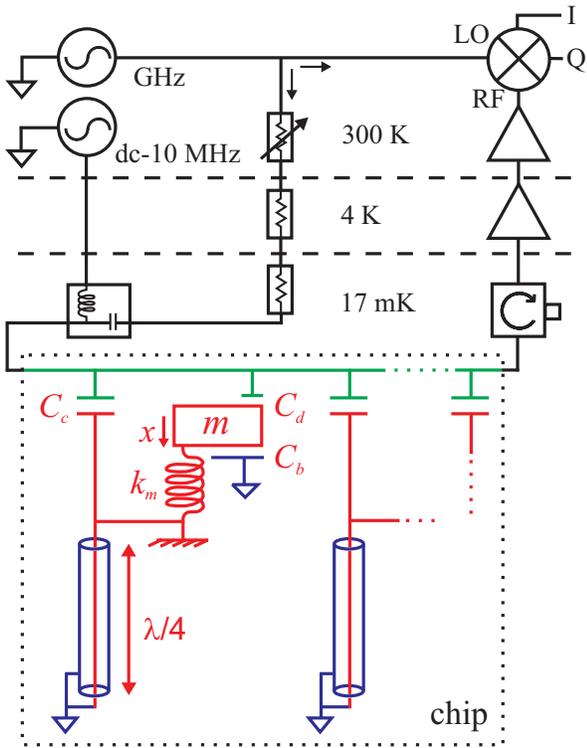} 
\caption{Measurement schematic.  Distributed microwave resonators (red and blue) with a line impedance of $Z_1=$70 $\Omega$
are capacitively coupled, $C_c$, to a feedline (green).  A nanomechanical beam (red) is coupled to each cavity via a
capacitance $C_b$ and to the feedline via $C_d$ used for an electrostatic drive.  The cavity coupling is characterized by
170 aF/$\mu$m and the drive coupling by 0.2 aF/$\mu$m. The beam motion is detected by measuring the phase shift of an
injected microwave signal.  This signal travels through the device and is then amplified first by a low-noise microwave HEMT
amplifier and further at room temperature before going to the rf port of an IQ mixer.} \label{schematic}
\end{center}
\end{figure}

\begin{figure} \begin{center}
\includegraphics[width=245pt]{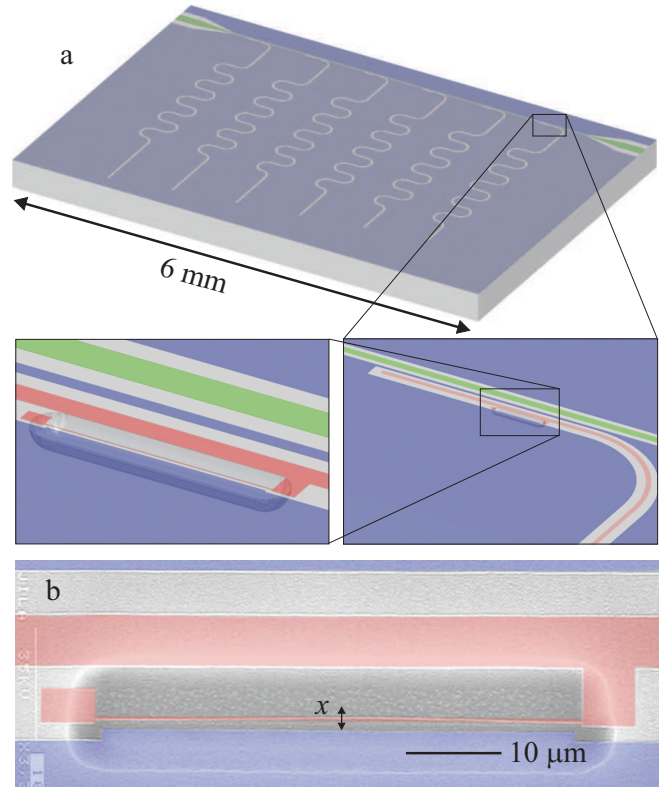}
\caption{(a) Drawing of our device showing frequency multiplexed  $\lambda / 4$ microwave cavities; the lines are meandered
to fit a quarter wave on the chip. The cavity lines are formed from 5 $\mu$m wide center conductors separated from the
ground plane by 10 $\mu$m slots.  The lower panels zoom into a capacitive elbow coupler and a nanomechanical beam with the
feedline shown in green, the ground plane in blue, and the center conductor in pink.    (b) False color scanning electron
microscope image of an embedded nanomechanical beam. This room temperature image shows a top view of the beam, which is
clamped on both ends and slightly bent due to compressive stress (see Methods).  An angled view of the same beam reveals
that it is also bent out of the plane at its center by 2.5 $\mu$m at room temperature. The area where the silicon was etched
to release the beam appears as the darker oblong region.} \label{pictures}
\end{center}
\end{figure}

The analogy between our microwave system and an optical cavity interferometer is quite rigorous; the Hamiltonian describing
both systems is \be H=\hbar \omega_0 (a^\dagger a+ \mbox{$\small \frac{1}{2}$}) + \hbar \omega_m (b^\dagger b +
\mbox{$\small \frac{1}{2}$}) - \hbar g a^\dagger a (b^\dagger+b)\delta x _{zp} \ee where the cavity and mechanical modes are
described respectively by the operators $a$ and $b$, $\omega_0$ and $\omega_m$ are the bare resonant cavity and mechanical
frequencies, and $g$ is the effect of the displacement $\hat{x}=(b^\dagger+b)\delta x _{zp}$ on the perturbed cavity
resonant frequency, $\omega_c$. In both cases the Heisenberg limit is enforced, as discussed above, by fluctuations in the
optical or microwave field, i.e. shot noise. Still there are important practical differences between optical experiments and
our microwave work. While the optical shot-noise limit is achieved routinely, due to the smaller photon energy of
microwaves, measurement of microwave fields is currently dominated by amplifier noise. Nonetheless, microwave amplifier
technology is progressing quickly, and in our experiments we use a commercially available HEMT amplifier that already
reaches a noise temperature of $k_b T_N/\hbar \omega_c=30$ ($T_N=7.5$ K). The small microwave photon energy also requires an
excellent force sensitivity to detect quantum backaction, but our experiments aim to accommodate this requirement by going
to the extremes of force sensitivity using floppy mechanical objects.

Our superconducting microwave cavities are formed from distributed transmission lines in a coplanar waveguide geometry and
are patterned using an aluminum thin film on a high-purity silicon substrate.   We use one-sided cavities that are shorted
on one end and coupled to a 50 $\Omega$ feedline on the other end (Fig. \ref{schematic} and Fig. \ref{pictures}(a)); the
cavity is overcoupled (dominated by external coupling not by internal losses) to minimize microwave power dissipation. We
study multiple cavities on a single chip by coupling six cavities to the feedline and address them individually via
frequency multiplexing \cite{Day2003}. The quarter wave resonances of our cavities are near $\omega_c = 2 \pi \times 5$ GHz.

The nanomechanical objects we embed in the cavity are thin, high aspect ratio beams of conducting Al clamped on both ends
(see Methods). We use a beam 50 $\mu$m long with a 100 nm by 130 nm cross-section (Fig. \ref{pictures}(b)).  The thin beam
gives us a small mass (an effective mass $m$ of 2 pg), while the length provides both good coupling to the microwave cavity
as well as a very small spring constant of a few mN/m. The beam is placed in the cavity such that the motion of its
fundamental flexural mode changes the capacitance between the cavity center conductor and the ground plane in a small
section of the cavity (Fig. \ref{pictures}). To maximize the coupling, the gap between the beam and the ground plane is as
small as is feasible (typically 1 $\mu$m), and the beam is embedded at a voltage antinode of the cavity standing wave. With
the beam at this position the cavity resonance frequency shifts according to $ \frac{1}{\omega_c} \frac{\partial
\omega_c}{\partial x} = -\frac{\partial C_b}{\partial x} 4 Z_1 \omega_c /2 \pi$ for a $\lambda/4$ cavity, where
$\frac{\partial C_b}{\partial x}$ is the effect of the beam motion on the cavity capacitance and note $-\frac{\partial
\omega_c}{\partial x}$ is the coupling $g$ of Eq. (1).

To detect nanomechanical motion with our microwave cavity interferometer we inject a microwave tone near a cavity resonance
and monitor the phase of the transmitted signal; this phase directly reflects the cavity resonance frequency and hence the
beam displacement as described above. Figure \ref{schematic} shows how we extract the phase (Q) and amplitude (I)
quadratures of the transmitted signal using a homodyne detection scheme.

\begin{figure} \begin{center}
\includegraphics[width=\linewidth]{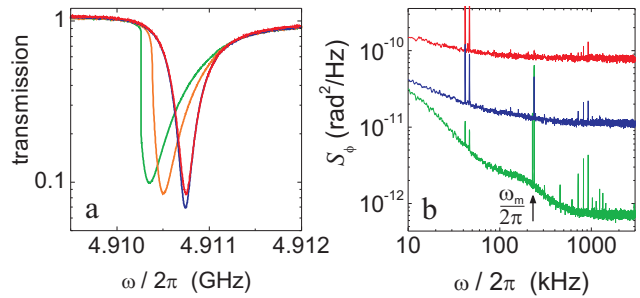}
\caption{(a) Power transmission past the cavity normalized to transmission off resonance.  The data are shown on a
logarithmic scale for incident powers of 1070 (green), 680 (orange),
 68 (blue), and 11 pW (red).  All microwave powers quoted in this work have a 3 dB systematic uncertainty.  At the highest microwave
powers the resonance becomes nonlinear and eventually bistable. At $P=68$ pW, a fit to the measured response reveals an
internal quality factor of $Q_{int} = 38,000$, an external quality factor of $Q_{ext} = 14,000$, and a total quality factor
of $Q = (Q_{int}^{-1}+Q_{ext}^{-1})^{-1}= 10,000$.  (b) Measurement of the double-sideband cavity phase noise at incident
powers of 1070 (green), 68 (blue), and 11 pW (red). Motion of the mechanical beam creates the tone indicated by the black
arrow; the other tones are caused by electronic interference.} \label{thephase}
\end{center}
\end{figure}

Figure \ref{thephase}(a) illustrates the microwave cavity resonant response.  Here we measure the relative transmission past
the cavity for a set of incident microwave powers $P$ at a dilution refrigerator temperature of 17 mK, far below the 1.2 K
critical temperature of Al.  For the highest microwave powers nonlinearities become significant as the current density
approaches the superconducting critical current density \cite{Dahm2002a}. At low microwave power the resonant behavior is
characterized by unity transmission off resonance and by a lorentzian response that dips to a value determined by the
intracavity losses compared to the feedline coupling.  Our imprecision in the nanomechanical beam position readout is
determined by fluctuations in the associated dispersive phase signal on resonance. In Fig. \ref{thephase}(b) we plot our
experimentally observed cavity phase fluctuations as the spectral density $S_\phi$.  At the highest response frequencies we
see a phase noise consistent with the HEMT amplifier noise, while at lower frequencies the phase noise is enhanced.  This
additional noise has been recently traced to two-level fluctuators in the silicon substrate \cite{Gao2007a}.

An important feature of our experiment is the ability to actuate the mechanical beam without applying large magnetic fields
that are incompatible with high-Q superconducting cavities \cite{Frunzio}. In our device we incorporate a small capacitive
coupling between the beam and the microwave feedline, which allows us to electrostatically drive the beam by coupling
low-frequency signals onto the feedline (Fig. \ref{schematic}). Using a bias-tee we introduce an ac signal near the
$\omega_m$ and a dc voltage resulting in an electrostatic force $ F_{el}(\omega) = V_{dc} V_{ac}(\omega) \frac{\partial
C_d}{\partial x} $ where $C_d$ is the drive capacitance between the feedline and the beam.  To assure that the effect of the
beam motion remains in the phase quadrature of the microwave signal, we design $C_d$ to be much smaller than $C_b$.

\begin{figure} \begin{center}
\includegraphics[width=\linewidth]{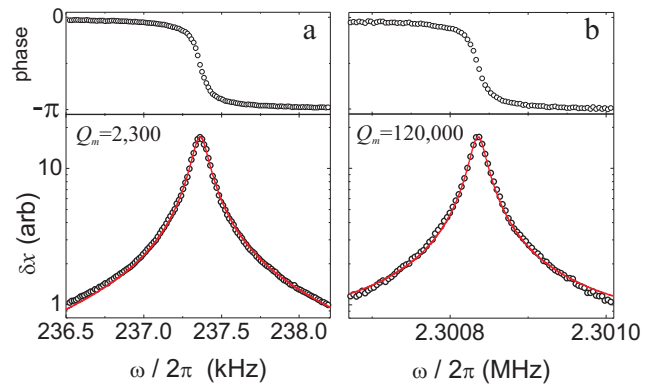}
\caption{(a) Resonant response of an aluminum nanomechanical beam to an electrostatic drive at $T_{frig}=$17 mK. This
experiment uses the beam shown in Fig. \ref{pictures}(b), and we find $Q_m=2,300$. (b) Response of a beam under tensile
stress.  Here the resonance is near 2 MHz and the quality factor is greatly enhanced to $Q_m=120,000$.  Note that as the
stress is varied the change in the dissipation, as reflected by the linewidth $\gamma_m = \omega_m/Q_m$, is not as dramatic
as the change in $Q_m$.  The red lines are the square root of lorentzian fits to the data.} \label{resonances}
\end{center}
\end{figure}

Figure \ref{resonances}(a) demonstrates nanomechanical displacement detection using our microwave cavity interferometer.
Here the beam motion we are measuring is the response of the beam illustrated in Fig. \ref{pictures}(b) to an electrostatic
drive. We see a clean response on a logarithmic scale, the expected $\pi$ phase shift, and good agreement with the
anticipated lorentzian response (red line) of our high-Q resonance. We measure a quality factor of $Q_m=2,300$ and find the
mechanical resonance at $\omega_m=2 \pi \times 240$ kHz; this frequency is near our expectation for a tension-free beam with
our geometry.

Figure \ref{resonances}(b) demonstrates the mechanical response we observe using a 50 $\mu$m long beam fabricated from an
aluminum film under tensile stress (see Methods).  The stress significantly increases $\omega_m$ to near $2\pi \times 2.3$
MHz and $Q_m$ to 120,000.  This quality factor is a surprisingly, yet pleasingly, large for a beam fabricated from an
amorphous metal and of this surface to volume ratio \cite{Ekinci2005a,Verbridge2006,Zwickl2007}.  When working with
mechanical objects with $\omega_m$ in the MHz regime we must take into account that the sidebands generated by the beam's
motion move outside the cavity bandwidth, $\gamma_c = \omega_c/Q=2\pi \times 490$ kHz.  In this so-called good-cavity limit,
to acquire the response seen in Fig. \ref{resonances}(b) we detune the injected microwave signal off-resonance by $\omega_m$
to place one sideband on the cavity resonance. Conversely if we decrease the beam tension to where $\omega_m \ll \gamma_c$
we are in the bad-cavity limit; the motion of the 240 kHz beam is mostly in this limit, and for this case we operate with
the injected microwave tone tuned to the cavity resonance.

\begin{figure} \begin{center}
\includegraphics[width=240pt]{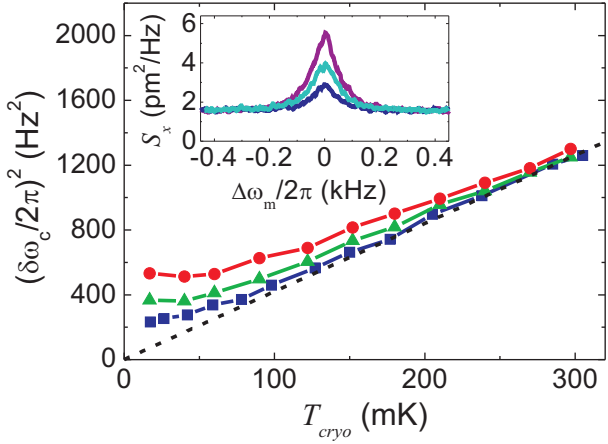}
\caption{Integrated mechanical beam fluctuations in units of cavity resonance frequency shift.  The three data sets
correspond to $P=4$ pW (blue squares), 27 pW (green triangles), and 68 pW (red circles). The dashed line shows the linear
fit described in the text from which we extract the coupling $g$. (Inset) Lorentzian response for dilution refrigerator
temperatures of 210 mK, 122 mK, and 40 mK at $P=11$ pW.} \label{thermal}
\end{center}
\end{figure}

By measuring the non-driven, thermomechanical motion of the beam, we can characterize the sensitivity and backaction of the
microwave measurement.  The expected thermally driven displacement fluctuations of our high-Q mechanical resonators at a
bath temperature $T$ is given by \be S_x(\Delta \omega_m) = \frac{1}{(m \omega_m \gamma_m)^2} \frac{4 m \gamma_m k_b T}{1+4
\Delta \omega_m^2 / \gamma_m^2} \ee  where $\Delta \omega_m=\omega-\omega_m$. The inset to Fig. \ref{thermal} shows
nondriven response at three different values of the dilution refrigerator temperature $T_{frig}$. (Here and in the remainder
of our experimental results we will be studying the 240 kHz mechanical beam discussed above.) The white noise background is
the imprecision $S_x^{im}$, while the height of the peak above the background describes the real fluctuations in the beam
position.  To understand the temperature-dependent response (and to calibrate the beam to cavity coupling) we can examine
the integrated signal under the lorentzian as a function of $T_{frig}$. Figure \ref{thermal} shows the integrated mechanical
motion in the experimental units of cavity resonance frequency shift $\delta \omega_c$ for a set of incident microwave
powers $P$ (see Methods).

In an ideal system, the integrated response should depend linearly on the temperature according to $  \delta \omega_c ^2 =
\frac{g^2 k_b}{m\omega_m^2} (T_{frig}+T_{ba}) $ where $T_{ba}=S_F^{ba}/4k_bm\gamma_m$ is the equivalent backaction
temperature. If we focus on lowest microwave power results (blue squares) and the highest dilution refrigerator temperatures
we see that the response is linear down to $\sim$100 mK.  Here linear fits reveal that the backaction is small compared to
relevant uncertainties, and we extract a coupling of $g = 2 \pi \times 1.16$ kHz/nm using points above 127 mK (dashed line).
This value of $g$ corresponds to a capacitance change of $\frac{\partial C_b}{\partial x}=170$ aF/$\mu$m, which is
consistent with our numerically calculated expectation.

At lower values of $T_{frig}$ and higher microwave powers (green triangles and red circles) the beam temperature decouples
from $T_{frig}$ leading to a saturation behavior.  The microwave power dependence suggests that the additional fluctuations
are related to microwave power dissipation.  However, by using a different cavity on the same chip as a crude thermometer,
we know that the dissipated power does not heat the entire chip above $T_{frig}$. The heating of the beam by the microwave
power must be a more local effect.

\begin{figure} \begin{center}
\includegraphics[width=\linewidth]{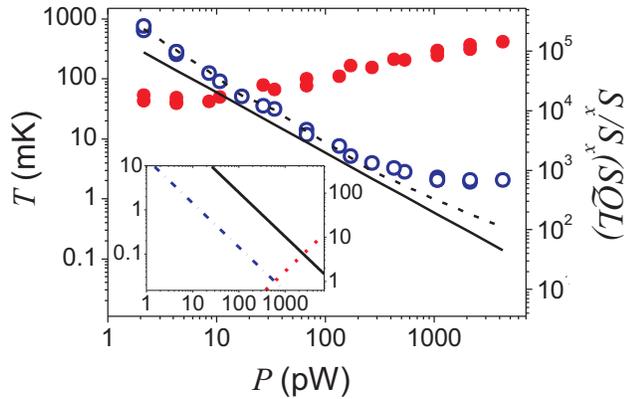}
\caption{Imprecision temperature (blue open circles) and saturation temperature (red circles) as a function of incident
microwave power.  The lines represent the expected imprecision due to our microwave amplifier; the dashed line includes the
loss of microwave power in the cavity, while the solid line represents the ideal expectation in the limit of a lossless
cavity.  On the right vertical axis we display the position uncertainty in units of $S_x(SQL)$. (Inset)  Over the same range
of power  we calculate the quantum-limit of displacement detection for {\it optimized parameters}.  We consider the ideal
case of a lossless one-sided cavity coupled to one port using the parameters quoted in the text.
 The lines correspond to the shot-noise limit (dash-dot line),
the quantum backaction (dotted line), and the imprecision due to a $T_N=$5 K voltage amplifier (solid line).} \label{power}
\end{center}
\end{figure}

Given the nonlinear dependence of the beam fluctuations on $T_{frig}$ we must be cautious in how we define and extract
$T_{ba}$ for our measurement.  We believe an honest metric is the equivalent temperature of the beam fluctuations at our
base temperature ($T_{frig}=17$ mK), which we will refer to as the saturation temperature $T_{sat}$.  This temperature is a
conservative upper limit on our ability to measure a backaction force since extrapolating $T_{ba}$ from only the highest $T$
points would yield a smaller value. In Fig. \ref{power} we plot $T_{sat}$ along with the imprecision temperature
$T_{im}=S_x^{im} m \omega_m^2 \gamma_m/4 k_b $ as a function of $P$.  At the lowest powers the imprecision dominates over
the beam fluctuations.  As the power is increased the imprecision drops linearly with power as expected, but at the highest
microwave powers it is enhanced by cavity phase noise to a value above our microwave amplifier noise floor (dashed line).

From the results in Fig. \ref{power} we can assess how close an approach we have made to the SQL.  The minimum uncertainty
in continuous position detection subject to the Heisenberg constraint $S_x^{im} S_F^{ba} = \hbar^2$ occurs at the point
where $S_x^{im} = S_x^{ba} =  S_F^{ba}/(m \omega_m \gamma_m)^2$.  Here the imprecision and the backaction both contribute
$S_x(SQL)=\hbar/m \omega_m \gamma_m$.  We can convert the imprecision and saturation temperatures in Fig. \ref{power} into a
position spectral density compared to this minimum value via $S_x/S_x(SQL) = 4 k_b T / \hbar \omega_m $.  This result is
shown on the right axis of Fig. \ref{power}.  Our limit on the imprecision {\it alone} at the highest microwave powers
corresponds to, in linear units, $30 \sqrt{S_x(SQL)}$.  The {\it total} minimum position uncertainty we achieve occurs at
$P=20$ pW and is given by $\sqrt{S_x^{tot}}=130 \sqrt{S_x^{tot}(SQL)}$, where to be explicit $S_x^{tot}$ here is
$S_x^{im}+S_x^{sat}$, which is an upper bound on our ability to measure $S_x^{im}+S_x^{ba}$.

We can also extract absolute values for the achieved sensitivity.  Our imprecision is limited to 200 $\rm fm/\sqrt{Hz}$ at
the highest microwave powers, which is a modest achievement compared to optical systems \cite{Arcizet2006}. On the other
hand our total force sensitivity, $\sqrt{S_F^{tot}} = \sqrt{4 k_b (T_{im}+T_{sat})m\gamma_m}$, is 3 $\rm aN/\sqrt{Hz}$ at
$P=20$ pW. This value is near the record mechanical force sensitivity of 0.8 $\rm aN/\sqrt{Hz}$ achieved using a fiberoptic
interferometer and a silicon cantilever \cite{Mamin2001}.

To closer approach the SQL with our microwave cavity interferometer the foremost task will be to decrease the dissipation
that likely leads to our observation of a finite $T_{sat}$; this dissipated power is determined by $Q_{int}$ compared to
$Q_{ext}$ and for our current work is 2 pW at $P=20$ pW. To decrease the dissipated power the nanomechanical beam processing
must be developed to be compatible with a very large $Q_{int}$. Another route to improvement is to increase $\omega_m$ which
will decrease thermal fluctuations and the dissipative force compared to quantum fluctuations, as well as decrease the
$\nu^{-1/2}$ cavity phase noise. However, to maintain the force sensitivity, an increase in the mechanical spring constant
should be accompanied by an increase in the beam to cavity coupling $g$. One option for increasing $g$ would be to decrease
the total cavity capacitance by operating at larger $\omega_c$ or by utilizing a higher-impedance microwave cavity or
lumped-element circuit.

It is instructive to assess what we could achieve in future experiments utilizing optimized, yet likely realizable,
parameters.  Consider a device described by $\omega_m= 2 \pi \times 2$ MHz, $m=2$ pg, $Q_{m}=100,000$, $\omega_c= 2 \pi
\times 12$ GHz, $Q_{int}\gg Q_{ext}=3,000$, and $g=2\pi \times 20$ kHz/nm. Further, assume we modify our geometry to measure
in reflection off of a single-sided cavity with a single port. For this more ideal geometry and a microwave probe at
$\omega_c$, the quantum-limited imprecision expected for detection via a classic square-law detector would be the shot-noise
limit \be S_x^{sn}=\frac{\hbar \omega_c (1+4(\omega_m/\gamma_c)^2)}{2(g/\omega_c)^2 P (4Q)^2}, \ee and correspondingly
$S_F^{ba}= \hbar^2 / S_{x}^{sn}$.  Our calculated expectation for $S_x^{sn}$ and $S_F^{ba}$ is shown in the inset to Fig.
\ref{power}.  We also include the noise of a possible HEMT amplifier (solid line).  This level could be improved by
incorporating better microwave amplifiers that could soon be available given recent interest in developing novel microwave
amplifiers near the quantum limit \cite{Yurke1989,Muck2001,Manuel}. However, even a quantum-limited voltage amplifier
measuring both field quadratures will result in an imprecision a factor of two above the shot-noise limit \cite{Caves1982a},
and to reach the SQL one must utilize an amplifier that detects only one quadrature.

Overall the results of the calculations in the inset to Fig. \ref{power} are promising; even with our current HEMT amplifier
the minimum uncertainty (assuming ideal backaction) would be a factor of $\sim$2.0 above the quantum limit in linear units. While
the equivalent backaction temperatures that must be measured are only a fraction of a mK, with precise measurements it is in
principle possible to extract a $T_{ba}$ that is much smaller than the bath temperature.  Most importantly, the shot-noise
limit and quantum backaction intersect at an achievable incident microwave power of 600 pW.

The novel coupled mechanical and microwave system we have demonstrated is not only promising as a detector, but also could
be adapted to cool nanomechanical resonators towards their ground state.  As first explored by Braginsky
\cite{Braginsky1977} dynamical backaction due to radiation pressure can lead to a passive cooling or heating of the
mechanical motion when the injected tone is detuned from the cavity resonance.  This effect has recently been observed in
optical cavities with micromechanical mirrors \cite{Arcizet2006b,Gigan2006,Kleckner2006,Schliesser2006,Thompson2007} and
radiofrequency circuits \cite{Brown2007}, and it has been suggested as a method of cooling and manipulating a beam coupled
to a transmission-line cavity \cite{Xue2007,Vitali2007}. Especially interesting is the possibility of passive cooling
utilizing our ability to access the good-cavity limit, as it is in this limit that one can in principle cool fully to the
mechanical ground state \cite{Marquardt2007a,WilsonRae2007}.  In addition to passive cooling, our ability to apply
electrostatic forces while reading out displacement with our microwave cavity interferometer makes us well-poised to
implement feedback cooling of our nanomechanical beams \cite{Poggio2007,Genes2008}.

\section{Methods}

The device is fabricated using a combination of electron-beam lithography and photolithography, and the beam and cavity are
formed from the same thermally evaporated aluminum film in a single liftoff process.  The beam is patterned directly on the
silicon substrate and suspended at the end of the process with an isotropic, dry silicon etch.  A relatively deep etch of 4
$\mu$m is typically required to release our mechanical beams with low spring constants. An insulating layer of SiO$_2$
underneath the rest of the pattern is used to protect the coplanar waveguide slots during the etch.

Our initial thermally evaporated aluminum film contains significant compressive stress.  To adjust the stress of the
aluminum beam we partially anneal the device at 150 - 350 $\rm{^o}$C in atmosphere before releasing the beam from the
substrate. The final stress of the beam at cryogenic temperatures is affected by the differential thermal coefficient of
expansion of silicon and amorphous aluminum.  We estimate that between room and cryogenic temperatures the aluminum film
shrinks by a few tenths of a percent compared to its clamping locations.  Hence, the beam of Fig. \ref{resonances}(a) has
significant compressive stress at room temperature (see Fig. \ref{pictures}(b)) but less compressive stress at $T_{frig}$;
the beam of Fig. \ref{resonances}(b) has little stress at room temperature but significant tensile stress at $T_{frig}$.

Since our IQ mixer (Marki IQ03076XP) has orthogonal outputs near 5 GHz, we can place all of the phase information in the Q
channel by rotating the phase of the signal into the LO of the mixer. The voltage fluctuations measured in the Q quadrature
$S_V^Q$ are then all that is required to extract the integrated cavity resonance frequency shift plotted in Fig.
\ref{thermal}.  The relationship between the cavity resonance frequency fluctuations and the voltage fluctuations is \be
S_{\omega_c} = \frac{\omega_c^2 (1 + 4(\omega_m/\gamma_c)^2)}{(2 Q)^2 V_0^2 (1 - S_{min})^2} S_V^Q \ee where $V_0$ is
voltage amplitude of the transmission off resonance and $S_{min}$ is the normalized transmission past the cavity on
resonance. The term $1 + 4(\omega_m/\gamma_c)^2$ accounts for filtering of the cavity response at $\omega_m$ and becomes 1
in the bad-cavity limit.  The integrated response in units of cavity resonance frequency shift is then given by $ \delta
\omega_c^2
 = S_{\omega_c}^0 \gamma_m/4$, where $S_{\omega_c}^0$ is the magnitude of the lorentzian response at $\Delta
\omega_m=0$. The $T$ dependence of the $Q$, $S_{min}$, and $Q_m$ must be taken into account in these conversions, but we
restrict our measurements to below 300 mK where the values change by $< 20 \%$.

In Fig. \ref{power} we extend our measurement into the regime where the cavity resonance becomes nonlinear (see Fig.
\ref{thephase}) and hence conversion between $S_V^Q$ and $S_{\omega_c}^Q$ becomes less straightforward to calculate.  To
extract this conversion at nonlinear microwave powers, we perform a separate calibration experiment in which we apply a
constant electrostatic drive and compare the beam response at high microwave power to the known response at low power.

\section{Acknowledgements}

We acknowledge support from the NSF Physics Frontier Center for AMO Physics and from the National Institute of Standards and
Technology.  C. A. R. acknowledges support from the Fannie and John Hertz Foundation.  We thank R. J. Schoelkopf, S. M.
Girvin, K. D. Irwin, D. Alchenberger, M. A. Castellanos-Beltran, and N. E. Flowers-Jacobs for enlightening conversations and
technical assistance.


\end{document}